\newcommand{\be}{\begin{equation}}\newcommand{\ee}{\end{equation}}
\newcommand{\bea}{\begin{eqnarray}}\newcommand{\eea}{\end{eqnarray}}
\newcommand{\nn}{\nonumber\\[6pt]}

\newcommand{\p}[1]{(\ref{#1})}

\newcommand{\bD}{\overline D}
\newcommand{\cD}{{\cal D}}
\newcommand{\cW}{{\cal W}}
\newcommand{\cU}{{\cal U}}
\newcommand{\cQ}{{\cal Q}}
\newcommand{\cbD}{{\overline{\cal D}}}
\newcommand{\cV}{{\cal V}}
\newcommand{\cbV}{{\overline{\cal V}}}

\newcommand{\bt}{{\bar\theta}}
\newcommand{\bxi}{{\bar\xi}}
\newcommand{\bpsi}{{\bar\psi}}

\newcommand{\beps}{{\bar\epsilon}}
\newcommand{\vt}{\vartheta}

\documentclass[12pt]{article}
\usepackage{amscd,amsmath,amssymb}

\begin{document}

\thispagestyle{empty}
\vspace{2cm}
\begin{flushright}
hep-th/0406015 \\
ITP-UH-12/04 \\
%LNF -???\\[5mm]
%\today \\[1cm]
\end{flushright}

\vspace{2cm}

\begin{center}
{\Large\bf ABC of N=8, d=1 supermultiplets }
\vspace{1cm}

{\large\bf S. Bellucci${\,}^{a}$, \ E. Ivanov${\,}^{b}$, \
S. Krivonos${\,}^{b}$, \ O. Lechtenfeld${\,}^{c}$ }

\vspace{1cm}
${}^a$ {\it INFN-Laboratori Nazionali di Frascati,
 C.P. 13, 00044 Frascati, Italy}

\vspace{0.1cm}
{\tt bellucci@lnf.infn.it}

\vspace{0.5cm}
${}^b$ {\it Bogoliubov  Laboratory of Theoretical Physics, JINR, 141980 Dubna,
Russia}

\vspace{0.1cm}
{\tt eivanov, krivonos@thsun1.jinr.ru}

\vspace{0.5cm}
${}^c$ {\it Institut f\"ur Theoretische Physik, Universit\"at Hannover,
 30167 Hannover, Germany}

\vspace{0.1cm}
{\tt lechtenf@itp.uni-hannover.de}
\end{center}

\vspace{2cm}

\begin{abstract}
\noindent
We construct a variety of off-shell $N{=}8, d{=}1$ supermultiplets
with finite numbers of component fields as direct sums of properly
constrained $N{=}4, d{=}1$ superfields.
We also show how these multiplets  can be described in
$N{=}8, d{=}1$ superspace where the whole amount of supersymmetry
is manifest. Some of these multiplets can be obtained by
dimensional reduction
{}from $N{=}2$ multiplets in $d{=}4$, whereas others cannot. We give examples
of invariant superfield actions for the multiplets constructed,
including $N{=}8$ superconformally invariant ones.
\end{abstract}
\vfill

\newpage
\setcounter{page}{1}
\section{Introduction}
Supersymmetric quantum mechanics (SQM) \cite{W} is relevant
to a vast bunch of phenomena associated with supersymmetric theories
in diverse dimensions and string theory
(see e.g.~\cite{Rev,rev2} and references therein).
In particular,
superconformal quantum mechanics (SCQM) \cite{SC1}--\cite{bikl}, \cite{rev2}
has profound implications in the
AdS/CFT \cite{ads1} and black holes arena
(see e.g. \cite{Azc}--\cite{BIGK}). It is also closely
related to the integrable Calogero-Moser-type systems \cite{cal1,BGK}.
Taking into account these and some other uses, the construction and analysis
of new SQM models (including SCQM models) is an urgent
and interesting task with many potentially important outcomes.

Most of the SQM models explored to date possess
$N{\leq}4, d{=}1$ supersymmetries. Much less is known
about higher-$N$ SQM models. Some of them were addressed many years
ago in the seminal paper \cite{Rit} within an on-shell Hamiltonian approach,
some others (with $N{=}8$) received attention lately in connection with
branes and black holes \cite{DE,Tbranes,BMZ}, and as effective theories
describing low-energy dynamics of BPS monopoles in $N{=}4$ super Yang-Mills
theory \cite{Monop}. $N{=}4$ and $N{=}8$ SQM models were also used to describe
$N{=}1$ and $N{=}2$ four-dimensional supersymmetric gauge theories in a
small spatial volume (see \cite{Smi1} and references therein).

The natural formalism for dealing with supersymmetric models is the
off-shell superfield approach. Thus for the construction of new SQM models
with extended $d{=}1$ supersymmetry, one needs, first of all, the
complete list of the corresponding off-shell $d{=}1$ supermultiplets
and the superfields which comprise these multiplets. One of
the peculiarities of $d{=}1$ supersymmetry is that some
of its off-shell multiplets cannot be obtained via direct
dimensional reduction from the multiplets of higher-$d$ supersymmetries
with the same number of spinorial charges. Another peculiarity is that
some on-shell multiplets
of the latter have {\it off-shell\/} $d{=}1$ counterparts. Taking into
account these and some other specific features of $d{=}1$ supersymmetry,
it is desirable to have convenient self-consistent methods of deducing
$d{=}1$ off-shell multiplets and relevant superfields directly within the
$d{=}1$ setting, without resorting to dimensional reduction.

One of such methods was proposed in \cite{leva2} and further
advanced in recent papers \cite{IKL,IKL2,bikl}. It is based
on nonlinear realizations of the
finite-dimensional superconformal groups in $d{=}1$ (see also \cite{Azc}).
The irreducible superfields representing one or another off-shell $d{=}1$
supermultiplet come out as the Goldstone superfields parametrizing one
or another coset manifold of the proper $d{=}1$ superconformal group.
The superfield irreducibility constraints naturally emerge as a part
of manifestly covariant inverse Higgs \cite{InvH} conditions on the relevant
Cartan superforms.
This method is advantageous in that it automatically specifies the
superconformal properties of the involved supermultiplets,  which
are of importance, e.g. when constructing the SCQM models associated with these
multiplets.\footnote{Sometimes, there is no need to proceed just from
superconformal algebras for deducing the correct superfield
irreducibility conditions; one can make use of much simpler superalgebras
which are related to the superconformal ones via a contraction and include
as subalgebras the corresponding $d{=}1$ Poincar\'e superalgebras
(see \cite{IKL2} for an instructive example).}

In \cite{IKL2}, the full set of off-shell
$N{=}4$ supermultiplets with {\bf 4} physical fermions (and a finite number
of auxiliary fields) was deduced, proceeding from
nonlinear realizations of the most general $N{=}4, d{=}1$ superconformal
group $D(2,1;\alpha)$ in its various coset superspaces.\footnote{Two
other $N{=}4, d{=}1$ superconformal groups, $SU(1,1\vert 2)$ and
$OSp(4^\star\vert 2)$,
can be treated as special cases of $D(2,1;\alpha)$, with $\alpha{=}0$ or ${-}1$
and $\alpha{=}1$, respectively \cite{FRS,VP}.}
Besides already known $N{=}4$ multiplets, some new off-shell multiplets were
found within this setting, viz. nonlinear `tensor'
and nonlinear chiral $N{=}4$ multiplets. The application of the same method
to the less studied case of $N{=}8, d{=}1$ supersymmetry was
initiated in \cite{bikl}. There we considered nonlinear realizations of
the $N{=}8, d{=}1$ superconformal group $OSp(4^\star\vert 4)$ in its two
different cosets and showed that two interesting $N{=}8, d{=}1$
multiplets, with off-shell field contents ({\bf 3, 8, 5})
and ({\bf 5, 8, 3}),
naturally come out as the corresponding Goldstone multiplets. We
constructed superconformally invariant actions for these multiplets in
$N{=}4, d{=}1$ superspace and revealed a surprising phenomenon which consists
in the following. The same $N{=}8, d{=}1$ supermultiplet admits a few
non-equivalent  splittings into pairs of irreducible off-shell $N{=}4, d{=}1$
multiplets, such that different $N{=}4$ superconformal subgroups of
$OSp(4^\star\vert 4)$, viz. $SU(1,1\vert 2)$ and $OSp(4^\star\vert 2)$,
are manifest for different splittings. Respectively, the
off-shell component action of the given $N{=}8$ multiplet in general
admits several different representations in terms of $N{=}4, d{=}1$ superfields.

The basic aim of the present paper is to give a superfield description
of all other linear off-shell $N{=}8, d{=}1$ supermultiplets with ${\bf 8}$
fermions, in both $N{=}8$ and $N{=}4$ superspaces.

Towards deriving an exhaustive list of off-shell $N{=}8$ supermultiplets and
the relevant constrained $N{=}8, d{=}1$ superfields, we could proceed in
the same way as in the case of $N{=}4$ supermultiplets
in \cite{IKL2}, i.e. by considering nonlinear realizations of all known
$N{=}8$ superconformal groups in their various cosets. However,
this task is more complicated as compared to the $N{=}4$ case,
in view of the existence of many non-equivalent
$N{=}8$ superconformal groups ($OSp(4^\star\vert 4)$, $OSp(8\vert 2)$, $F(4)$
and $SU(1, 1\vert 4)$, see e.g. \cite{VP}), with numerous coset manifolds.

In order to avoid these complications, we take advantage of two fortunate
circumstances.  Firstly, from the results of \cite{GR,PaTo} it
follows that the field contents of {\it linear\/} off-shell
multiplets of $N{=}8, d{=}1$ supersymmetry with {\bf 8} physical fermions
range from \break ({\bf 8, 8, 0}) to ({\bf 0, 8, 8}), with the intermediate
multiplets corresponding to all possible divisions of ${\bf 8}$
bosonic fields into physical and auxiliary ones. Thus we are aware of
the full list of such multiplets independently of the issue
of their interpretation as the Goldstone ones parametrizing the
proper superconformal cosets. Besides the two extreme possibilities,
\be
(\mbox{a})\;({\bf 8, 8, 0})\,, \quad (\mbox{b})\;({\bf 0, 8, 8})\,,
\label{Extr}
\ee
we have seven further ones,
\bea
&&
(\mbox{c})\;({\bf 7, 8, 1})\,, \;\;(\mbox{d})\;({\bf 6, 8, 2})\,, \;\;
(\mbox{e})\;({\bf 5, 8, 3})\,, \;\; (\mbox{f})\;({\bf 4, 8, 4})\,, \nn
&& \qquad\quad
(\mbox{g})\;({\bf 3, 8, 5})\,, \;\; (\mbox{h})\;({\bf 2, 8, 6})\,, \;\;
(\mbox{i})\;({\bf 1, 8, 7})\,. \label{Interm}
\eea
The superfield description of the multiplets (\ref{Interm}e) and
(\ref{Interm}g) was given in \cite{bikl}. The superfield description
of the remaining multiplets is just the subject of the present paper.

The second circumstance allowing us to advance without resorting to
the nonlinear realizations techniques is the aforesaid
existence of various splittings of $N{=}8$ multiplets into pairs of irreducible
$N{=}4$ supermultiplets. We know how to represent the latter in terms
of constrained $N{=}4$ superfields, so it proves to be a matter of simple
algebra to guess the form of the four extra supersymmetries mixing
the $N{=}4$ superfields inside each pair and extending the manifest
$N{=}4$ supersymmetry to $N{=}8$. After fixing such pairs, it is again
rather easy to embed them into the appropriately constrained $N{=}8, d{=}1$
superfields. Beyond this, one can wonder how to reproduce these
superfields within the superconformal coset techniques, as it was done for
the multiplets (\ref{Interm}e) and (\ref{Interm}g) in \cite{bikl}. We hope
to return to the analysis of this important question elsewhere.

The paper is organized as follows. In the preparatory Section~2
we list all the known
finite-dimensional off-shell multiplets of $N{=}4, d{=}1$ Poincar\'e
supersymmetry and recall their $N{=}4$ superfield formulations,
with the corresponding constraints \cite{IKL2}.
In Section~3 we sketch the construction of the $N{=}8$, $d{=}1$ superspace most
suitable for our
purposes. Section~4 is split into nine subsections related
to the nine different $N{=}8$ off--shell multiplets we are
considering. There we describe the structure of the multiplets in
terms of $N{=}4$ and $N{=}8$ superfields.
In particular, we construct the corresponding irreducibility constraints in
$N{=}4$ superspace, as
well as the implicit $N{=}4$ Poincar\'e supersymmetry transformations,
completing the manifest
ones to the full $N{=}8$ supersymmetry. Invariant free actions are also
constructed for all $N{=}8$
supermultiplets, and interaction terms are given for a few of them.

\setcounter{equation}0
\section{N=4, d=1 supermultiplets}
In this Section, based upon Ref.~\cite{IKL2}, we tabulate all possible
off-shell multiplets of $N{=}4, d{=}1$ supersymmetry
with finite number of component fields
and the corresponding
superfields as Goldstone multiplets of the nonlinearly realized
most general $N{=}4, d{=}1$ superconformal group $D(2,1;\alpha)$ (for the generic
$\alpha $ or for some special values of this parameter).

We use the following definition  of the $N{=}4, d{=}1$
spinor derivatives
\be
D^i=\frac{\partial}{\partial\theta_i}+i\bt^i \partial_t\; , \;
\bD_i=\frac{\partial}{\partial\bt^i}+i\theta_i \partial_t\; , \;
\left\{ D^i, \bD_j\right\}= 2i \delta^i_j \partial_t \; .
\ee
Sometimes it will be more convenient to use the quartet form of
the same derivatives:
\be
D^{ia} =
\frac{\partial}{\partial\theta_{ia}} + i\theta^{ia}\partial_t \equiv
(D^i, \bar D^i)\,,
\quad \theta_{ia}\equiv (\theta_i, -\bar\theta_i) \,. \label{def22}
\ee
Respectively, there are two equivalent parametrizations of the $N{=}4, d{=}1$ superspace
\be
\mathbb{R}^{(1|4)} = (t, \,\theta_i,\, \bar\theta^k) = (t,\, \theta^{ia})\,.\label{SS0}
\ee

In the Table below, besides the field content of multiplets and the
corresponding superfields, we also indicate to which coset of the
internal R-symmetry $SU(2)$ of the $N{=}4, d{=}1$ Poincar\'e
superalgebra the physical bosonic fields (or a subset of them)
can belong and whether one of these fields can be identified with
the Goldstone coset field (dilaton) associated with the dilatation
generator $\subset D(2,1;\alpha)$. On all these multiplets, except
for the chiral one, the supergroup  $D(2,1;\alpha)$ can be realized
at any value of the parameter $\alpha$. The $N{=}4, d{=}1$ chirality is
compatible only with the choice of the supergroup $SU(1,1\vert 2)$
as the superconformal one.
\vspace{0.5cm}

\begin{center}
{\bf N=4 supermultiplets }
%\end{center}
\vspace{0.4cm}

\begin{tabular}{|l|c|c|c|c|c|}
\hline
multiplet & content & R symm. & dilaton & $\alpha$ & SF \\
\hline
`old tensor' & ({\bf 1, 4, 3}) &  -- & yes & any & $u$  \\
  chiral & ({\bf 2, 4, 2}) &  c.charge & yes & $0, -1$ & $\phi, \bar\phi$  \\
nl. chiral& ({\bf 2, 4, 2}) &  $su(2)/u(1)$ & no & any & $\Lambda,
\bar\Lambda$  \\
tensor & ({\bf 3, 4, 1}) & $su(2)/u(1)$ & yes & any & $V^{ij}$  \\
nonlinear & ({\bf 3, 4, 1}) & $su(2)$ & no & any & $N^{ia}$  \\
hyper & ({\bf 4, 4, 0}) &  $su(2)$ & yes & any & $q^i,{\bar q}_i$\\
ferm. hyper & ({\bf 0, 4, 4}) &  --  & no & any & $\xi^a, \bxi_a$\\
\hline
\end{tabular}
\end{center}
\vspace{0.5cm}

\noindent The constraints read as follows:
\begin{itemize}
\item {`Old Tensor'}: $D^iD_i u = \bD{}^i\bD_i u =0$ \\
        {(Note: $\left[ D^i, \bD_i\right]u= const$; generic $\alpha$ is compatible only with $const = 0$, otherwise
$\alpha = 0$ or $-1$)}
\item {Chiral}: $D^i\phi=\bD_i\bar\phi=0 $ \\
        {(Note: it exists only for $[D(2,1;-1)$ or $D(2,1; 0)] \sim su(1,1|2)\times su(2))$}
\item {Nonlinear Chiral}: $D^1 \Lambda = -\Lambda D^2\Lambda, \;
\bD_2\Lambda=\Lambda\bD_1\Lambda $ (and c.c.)
\item {Tensor}: $D^{(i}V^{jk)}=\bD{}^{(i}V^{jk)}=0$
\item {Nonlinear}: $N^{a(i}D^j N^{k)}_a=N^{a(i}\bD{}^j N^{k)}_a=0\,,$\quad
$N^{ai}N_{ai}=2$
\item {Hypermultiplet}: $D^{(i} q^{j)}=\bD{}^{(i}q^{j)}=0$ (and c.c.)
\item {Fermionic Hypermultiplet}: $D^{(i} \xi^{j)}=\bD{}^{(i}\xi^{j)}=0$ (and c.c.)
\end{itemize}

\setcounter{equation}0
\section{N=8, d=1 superspace}
The maximal automorphism group of $N{=}8, d{=}1$ super Poincar\'e algebra
(without central charges) is $SO(8)$ and so eight real Grassmann coordinates
of $N{=}8, d{=}1$ superspace $\mathbb{R}^{(1|8)}$ can be arranged into one of three 8-dimensional
real irreps of $SO(8)$. The constraints defining the irreducible
$N{=}8$ supermultiplets in general break this $SO(8)$ symmetry. So,
it is preferable to split the
8 coordinates into two real quartets
\be\label{def1}
\mathbb{R}^{(1|8)} = (t, \,\theta_{ia},\, \vt_{\alpha A})\,,\qquad\quad
\overline{ \left( \theta_{ia}\right)}=\theta^{ia},\;
\overline{ \left( \vt_{\alpha A}\right)}=\vt^{\alpha A},
\quad i,a,\alpha, A= 1,2\,,
\ee
in terms of which only four commuting automorphism $SU(2)$ groups will be explicit.
The further symmetry breaking can be understood as identification of some of these $SU(2)$, while
extra symmetries, if exist, mix different $SU(2)$ indices.
The corresponding covariant derivatives are defined by
\be\label{def2}
D^{ia}=\frac{\partial}{\partial\theta_{ia}}+i\theta^{ia} \partial_t\; , \;
\nabla^{\alpha A}=\frac{\partial}{\partial\vt_{\alpha A}}
+i\vt^{\alpha A} \partial_t\;.
\ee
By construction, they obey the algebra:\footnote{We use the following convention for the skew-symmetric
tensor $\epsilon$: $\;\epsilon_{ij} \epsilon^{jk}=\delta_i^k \;,
\quad \epsilon_{12} = \epsilon^{21} =1 \;$.}
\be\label{def3}
\left\{ D^{ia}, D^{jb}\right\} =2i \epsilon^{ij}\epsilon^{ab}\partial_t\,,\quad
\left\{ \nabla^{\alpha A}, \nabla^{\beta B}\right\}
=2i \epsilon^{\alpha\beta}\epsilon^{AB}\partial_t \; .
\ee

Note that the direct dimensional reduction from $N{=}2, d{=}4$ superspace
to $d{=}1$ yields a superspace with two manifest $SU(2)$ automorphism groups
realized on the Grassmann coordinates: one of them is just the R-symmetry of
$N{=}2, d{=}4$ Poincar\'e superalgebra and  the other originates from the
$d{=}4$ Lorentz group $SL(2,C)$. These two $SU(2)$ correspond to the appropriate
identifying of the doublet indices of two groups $SU(2)$ in the
above product of four such groups. In what follows we shall not be bound to any sort
of dimensional reduction; the differential constraints on the relevant
$N{=}8, d{=}1$ superfields will be defined directly within the $d{=}1$ setting, in
terms of the spinor covariant derivatives \p{def2}.

\setcounter{equation}0
\section{N=8, d=1 supermultiplets}
As mentioned in Introduction, our real strategy of deducing a superfield
description of the $N{=}8, d{=}1$ supermultiplets \p{Extr}, \p{Interm}
consisted in selecting the appropriate pair of the constrained $N{=}4, d{=}1$
superfields from those listed in Section~2 and then in guessing the constrained
$N{=}8$ superfield which accommodate this pair in a most economical
way. In the Subsections below, just to make the presentation more
coherent, we turn the argument around and start with postulating the $N{=}8, d{=}1$
constraints. The $N{=}4$ superfield formulations will be deduced from the $N{=}8$ ones.

\subsection{Supermultiplet ({\bf 0, 8, 8})}

The off-shell $N{=}8, d{=}1$ supermultiplet ({\bf 0, 8, 8}) is carried out by two real
fermionic $N{=}8$ superfields $\Psi^{aA}, \Xi^{i\alpha}$ subjected to the following
constraints:
\bea
&& D^{(i a}\Xi^{j)}_{\alpha}=0,\; D^{i(a}\Psi^{b)}_{A}=0,\quad
\nabla^{(\alpha A}\Xi^{\beta)}_{i}=0,\;
\nabla^{\alpha( A}\Psi^{B)}_{a}=0, \label{1con1} \\
&& \nabla^{\alpha A}\Psi_A^a=D^{ia}\Xi_i^\alpha,\quad
\nabla^{\alpha A}\Xi_\alpha^i=-D^{ia}\Psi_a^A\;. \label{1con2}
\eea

In order to understand the structure of this supermultiplet in terms of $N{=}4$
superfields and to prove that the above constrained $N{=}8$ superfields
indeed carry the irreducible off-shell field content ({\bf 0, 8, 8}),
we proceed as follows. As the first step, let us single out
the $N{=}4$ subspace in the $N{=}8$ superspace $\mathbb{R}^{(1|8)}$ as the set of coordinates
\be
\mathbb{R}^{(1|4)} = \left( t, \theta_{ia} \right) \subset \mathbb{R}^{(1|8)}, \label{ss1}
\ee
and expand the $N{=}8$ superfields over the
extra Grassmann coordinate $\vartheta_{\alpha A}$. Then we observe
that the constraints \p{1con2} imply that the spinor derivatives
of all involved superfields with respect to $\vartheta_{\alpha A}$
can be expressed in terms of
spinor derivatives with respect to $\theta_{i a}\,$. This means that
the only essential $N{=}4$ superfield components of $\Psi^{aA}$ and
$\Xi^{i\alpha}$ in their $\vartheta$-expansion are the first ones
\be\label{1comp1}
\psi^{aA} \equiv \Psi^{aA}|_{\vt=0}\;,\quad \xi^{i\alpha} \equiv
\Xi^{i\alpha}|_{\vt=0}\,.
\ee

These fermionic $N{=}4$ superfields are subjected, in virtue of
eqs. \p{1con1} and \p{1con2}, to the irreducibility constraints in $N{=}4$
superspace
\be\label{1con3}
D^{a(i} \xi^{j)\alpha}=0,\;
D^{i(a} \psi^{b)A}=0\,.
\ee
As follows from the list of Section~2, these superfields are just
two fermionic $N{=}4$ hypermultiplets, each carrying ({\bf 0, 4, 4})
independent component fields. So, being combined together, they accommodate
the whole off-shell component content of the  $N{=}8$
multiplet ({\bf 0, 8, 8}), which proves that the $N{=}8$ constraints
\p{1con1}, \p{1con2} are the true choice.

Thus, from the $N{=}4$ superspace perspective, the
$N{=}8$ supermultiplet ({\bf 0, 8, 8}) amounts to the sum of two $N{=}4, d{=}1$
fermionic hypermultiplets  with the
off-shell component content ({\bf 0, 4, 4}) $\oplus$ ({\bf 0, 4, 4}).

The transformations of the implicit $N{=}4$ Poincar\'e supersymmetry,
completing the manifest one to the full $N{=}8$ supersymmetry, have the
following form in terms of the $N{=}4$ superfields defined above:
\be\label{1tr1}
\delta\psi^{aA}=\frac{1}{2}\eta^{A\alpha} D^{ia}\xi_{i\alpha},\quad
\delta \xi^{i\alpha}=-\frac{1}{2}\eta^\alpha_A D^i_a \psi^{aA}.
\ee

The compatibility of \p{1tr1} with the constraints \p{1con3}
is verified  using the following corollaries of these constraints:
\be\label{1corol}
D^{ia}D^{jb} \psi^{cA}=2i\epsilon^{ij}\epsilon^{cb}{\dot{\psi}}{}^{aA},\;
D^{ia}D^{jb} \xi^{k\alpha}=2i\epsilon^{ab}\epsilon^{kj}{\dot{\xi}}{}^{i\alpha}.
\ee

The invariant free action can be written as \be\label{1action}
S=\int dt d^4\theta \left[ \theta^{ia}\theta_i^b \psi_a^A
\psi_{bA}+\theta^{ia}\theta_a^j \xi^\alpha_i \xi_{j\alpha}\right].
\ee

Because of the presence of explicit theta's in the action
\p{1action}, the latter
is not manifestly invariant even with respect to the manifest $N{=}4$
supersymmetry. Nevertheless, one can check that \p{1action}
is invariant under this supersymmetry which is realized on the
superfields as
\be\label{1transf3}
\delta^* \psi^{aA}=-\varepsilon_{jb}Q^{jb} \psi^{aA},\;
\delta^* \xi^{i\alpha}=-\varepsilon_{jb}Q^{jb} \xi^{i\alpha},
\ee
where
\be
Q^{ia}=\frac{\partial}{\partial\theta_{ia}}-i\theta^{ia} \partial_t\;,
\ee
$\varepsilon_{ia}$ is the supertranslation parameter and
${}^*$ means the `active' variation (taken at a fixed point of
the $N{=}4$ superspace). Actually, the action \p{1action} can be given
the manifestly $N{=}4$ supersymmetric form as a sum of two integrals
over the appropriate harmonic analytic subspaces of $N{=}4, d{=}1$ superspace
(see \cite{IL}). Below we will employ this equivalent harmonic superspace form of
the off-shell action of the fermionic hypermultiplet (Subsections 4.2 and 4.5), as well as
one more its form, as an integral over the chiral $N{=}4, d{=}1$ superspace (Subsection 4.3).

\subsection{Supermultiplet ({\bf 1, 8, 7})}
This supermultiplet can be described by a single scalar $N{=}8$ superfield
$\cal U$ which obeys the following irreducibility conditions:
\bea
&& D^{ia}D_a^j{\cal U}=-\nabla^{\alpha j}\nabla_{\alpha}^i{\cal U},
\label{2con1a} \\
&& \nabla^{(\alpha i}\nabla^{\beta)j}{\cal U}=0,\quad
D^{i(a}D^{j b)}{\cal U} =0 \;. \label{2con1b}
\eea
Let us note that the constraints \p{2con1a} reduce the manifest
R-symmetry to $[SU(2)]^3$ due to the identification
of the indices $i$ and $A$ of the covariant derivatives
$D^{ia}$ and $\nabla^{\alpha A}$.

This supermultiplet possesses a unique decomposition
into the pair of the $N{=}4$ supermultiplets as
({\bf 1, 8, 7}) = ({\bf 1, 4, 3}) $\oplus$ ({\bf 0, 4, 4}). The corresponding
$N{=}4$ superfield projections  can be defined as
\be\label{2defcomp}
u={\cal U}|_{\vt=0} ,\quad \psi^{i\alpha}=\nabla^{\alpha i}{\cal U}|_{\vt=0}\;,
\ee
and they obey the standard constraints
\be\label{2con2}
D^{(i a}\psi^{j)\alpha}=0\;, \quad D^{i(a}D^{j b)} u=  0\;.
\ee
The second constraint directly follows from \p{2con1b}, while the first one is implied
by the relation
\be\label{2con1c}
\frac{\partial}{\partial t} D^{(i}_a\nabla^{j)}_{\alpha}{\cal U} =0\,,
\ee
which can be proven by applying the differential operator $D^{kb}\nabla^{\beta l}$
to the $N{=}8$ superfield constraint \p{2con1a} and making use of the algebra of covariant derivatives.

The  additional implicit $N{=}4$ supersymmetry is realized on these
$N{=}4$ superfields as follows:
\be\label{2tr}
\delta u=-\eta_{i\alpha}\psi^{i\alpha},\quad \delta \psi^{i\alpha}
=-\frac{1}{2}\eta^\alpha_j D^{ia} D^j_a u \;.
\ee
The simplest way to deal with the action for this supermultiplet
is to use harmonic superspace \cite{{harm},{book},{IL}},
at least for the spinor superfields $\psi^{i\alpha}$.

We use the definitions and conventions of Ref. \cite{IL}. The harmonic
variables parametrizing  the coset $SU(2)_R/U(1)_R$
are defined by the relations
\be\label{h1}
u^{+i}u^-_i = 1 \quad \Leftrightarrow\quad
u^+_i u^-_j - u^+_j u^-_i = \epsilon_{ij} \;, \;\;
\overline{(u^{+i})} = u^-_i\,.
\ee
The harmonic projections of
$\psi^{i\alpha}=\left\{\psi^i,\bar\psi{}^i\right\}$ are defined by
\be
\psi^{+}=\psi^{i} u^+_i,\;  \bpsi^{+}=\bpsi^{i} u^+_i,\;\label{h2}
\ee
and the constraints \p{2con2} are rewritten as
\bea
&& D^+ D^-u=\bD^+ \bD^-u=0,\quad D^+\bD^- u + \bD^+D^- u=0 \;,\label{h3a} \\
&& D^{+} \psi^{+}=\bD^{+}\psi^+=0\;, \; D^{++}\psi^{+}=D^{++}\bpsi^+= 0\;.
\label{h3b}
\eea
Here $D^{i\alpha}= \left(D^i,\bD^i \right)$ and
$D^{\pm} =D^{i} u^\pm_i, \bD^\pm=\bD^{i} u^\pm_i$,
$D^{\pm\pm} = u^{\pm i}\partial/\partial u^{\mp i}$ (in the central
basis of the harmonic superspace), with $D^{ia} $ given in \p{def22}.
The relations \p{h3b} imply that
$\psi^{+},\bpsi^+$ are analytic harmonic $N{=}4, d{=}1$ superfields
living on the analytic subspace $(\zeta, u^\pm_i)
\equiv (t_A, \theta^+, \bar\theta^+, u^\pm_i)$ which is closed under
$N{=}4$ supersymmetry. In this setting, the transformations of the hidden
$N{=}4$ supersymmetry
\p{2tr} are rewritten as
\bea\label{h4}
&& \delta^*\psi^+=\frac{1}{2}\left[ \eta^+ D^+\bD{}^-
- \eta^+\bD{}^+ D^- -2 \eta^- D^+ \bD{}^+ \right] u \;,    \nn
&&\delta^*\bpsi{}^+=\frac{1}{2}\left[ \bar\eta^+ D^+\bD{}^-
- \bar\eta^+\bD{}^+ D^- -2 \bar\eta^- D^+ \bD{}^+ \right] u \;, \nn
&& \delta^*u=-\bar\eta^-\psi^++\bar\eta^+\psi^-+\eta^-\bar\psi{}^+
-\eta^+\bar\psi{}^-\;.
\eea

The action is given by
\be\label{h5}
S= \frac{1}{2}\int dt d^4\theta \,u^2
+\int du d\zeta^{--}\psi^+\bpsi{}^+
\ee
where $du d\zeta^{--}= du dt_Ad\theta^+d\bar\theta^+ $
is the measure of integration over the analytic
superspace.
The action \p{h5} is manifestly $N{=}4$ supersymmetric since
it is written in terms of $N{=}4$ superfields.
However, its invariance with respect to the hidden $N{=}4$ supersymmetry
\p{h4} must be explicitly checked.
The variation of the first term in \p{h5} can be represented as
(we explicitly write only the terms involving the parameters $\eta^\pm$)
\be\label{h6a}
\delta\, \int dt d^4\theta \frac{u^2}{2}
= \int dt d^4\theta du \left(\eta^-\bpsi{}^+ - \eta^+\bpsi{}^-\right)u =
2\int dt d^4\theta du \eta^-\bpsi{}^+ u \,,
\ee
while the variation of the second term reads
\be\label{h6b}\
\delta \int du d\zeta^{--}\psi^+\bpsi{}^+
=-2 \int du d\zeta^{--} D^+\bD{}^+\left( \eta^- \bpsi{}^+ u \right).
\ee
Keeping in mind that
\be
\int du d\zeta^{--} D^+\bD{}^+ \equiv \int dt d^4\theta du \,,
\ee
we see that the action \p{h5} is indeed invariant under $N{=}8$ supersymmetry.

\subsection{Supermultiplet ({\bf 2, 8, 6})}
The $N{=}8$ superfield formulation of  this supermultiplet involves
two scalar bosonic superfields ${\cal U}, \Phi$ obeying
the constraints
\bea
&& \nabla^{(a i}\nabla^{b)j}{\cal U}=0, \quad
\nabla^{a(i}\nabla^{b j)}\Phi=0 , \label{31} \\
&& \nabla^{ai}{\cal U}=D^{ia}\Phi, \quad \nabla^{ai}\Phi=-D^{ia}{\cal U}
\label{32}
\eea
where we have identified indices $i$ and $A$, $a$ and $\alpha$ of the
covariant derivatives, thus retaining only two manifest SU(2) automorphism
groups.  From \p{31}, \p{32} some useful corollaries follow:
\bea
&& D^{ia}D^j_a{\cal U}+\nabla^{aj}\nabla^i_a{\cal U}=0,
\quad D^{i(a}D^{j b)}{\cal U}=0, \label{33} \\
&& D^{ia}D_i^b \Phi+\nabla^{b i}\nabla_i^a \Phi=0, \quad
D^{(i a}D^{j)b}\Phi=0.\label{34}
\eea
Comparing \p{33}, \p{34} and \p{31} with \p{2con1a}, \p{2con1b},
we observe that the $N{=}8$ supermultiplet with the field content
({\bf 2, 8, 6}) can be obtained by combining  two ({\bf 1, 8, 7})
supermultiplets and imposing the additional
relations \p{32} on the corresponding $N{=}8$ superfields.

In order to construct the invariant actions
and to prove that the above $N{=}8$ constraints indeed yield
the multiplet ({\bf 2, 8, 6}), we should reveal the structure
of this supermultiplet in terms of $N{=}4$ superfields,
as we did in the previous cases. However, in the case at hand,
we have two different choices to split the ({\bf 2, 8, 6})
supermultiplet
\begin{itemize}
\item {\bf 1. } ({\bf 2, 8, 6}) = ({\bf 1, 4, 3}) $\oplus$ ({\bf 1, 4, 3})
\item {\bf 2. } ({\bf 2, 8, 6}) = ({\bf 2, 4, 2}) $\oplus$ ({\bf 0, 4, 4})
\end{itemize}
As already mentioned, the possibility to have a few  different
off-shell $N{=}4$ decompositions of the same $N{=}8$ multiplet
has been observed in our earlier paper \cite{bikl}.
It is related to different choices of the manifest $N{=}4$
supersymmetries as subgroups of the $N{=}8$ super Poincar\'e group.
We shall treat both options. \vspace{0.5cm}

\noindent{\bf 1. } ({\bf 2, 8, 6}) = ({\bf 1, 4, 3}) $\oplus$ ({\bf 1, 4, 3})\\

In order to describe the $N{=}8$ ({\bf 2, 8, 6}) multiplet in terms of
$N{=}4$ superfields we should choose the appropriate $N{=}4$ superspace.
The first (evident) possibility is to choose the $N{=}4$ superspace
with the coordinates $( t, \theta_{ia} )$ i.e. the same as in
Subsections 4.1 and 4.2 (see eq. \p{ss1}). In this superspace
one  $N{=}4$ Poincar\'e supergroup is naturally realized, while the second
one mixes two irreducible $N{=}4$
superfields which comprise the  $N{=}8$ ({\bf 2, 8, 6}) supermultiplet
in question. Expanding the $N{=}8$ superfields ${\cal U},\Phi$
in $\vt^{ia}$,  one finds that the constraints \p{31}, \p{32} leave
in ${\cal U}$ and $\Phi$ as independent $N{=}4$ projections
only those of zeroth order in $\vt^{ia}$
\be\label{4def1}
\left. u={\cal U}\right|_{\vt_{i\alpha}=0}\,,\quad \left.
\phi=\Phi\right|_{\vt_{i\alpha}=0}\,.
\ee
Each $N{=}4$ superfield proves to be subjected, in virtue
of \p{31}, \p{32}, to the additional constraint:
\be\label{4aconn4}
D^{i(a}D^{j b)}u=0\,, \quad D^{(i a}D^{j)b}\phi=0\,.
\ee
Thus we conclude that our $N{=}8$  multiplet ${\cal U}, \Phi$, when
rewritten in terms of $N{=}4$ superfields, amounts to a direct sum
of two  $N{=}4$ multiplets $u$ and $\phi$, both having the same
off-shell field contents ({\bf 1, 4, 3}).

The transformations of the implicit $N{=}4$ Poincar\'e supersymmetry
completing the manifest
one to the full $N{=}8$ Poincar\'e supersymmetry have the following form
in terms of these $N{=}4$ superfields:
\be\label{4transf1}
\delta^* u= - \eta_{ia}D^{ia}\phi, \quad \delta^* \phi= \eta_{ia}D^{ia}u \;.
\ee

It is rather easy to construct the action in terms of $N{=}4$ superfields
$u$ and $\phi$, such that it is  invariant with respect
to the implicit $N{=}4$ supersymmetry \p{4transf1}. The generic action has the form
\be\label{4aaction1}
S= \int dt d^4\theta {\cal F}(u,\phi)\;,
\ee
where the function ${\cal F}$ obeys the Laplace equation
\be\label{4alaplace}
{\cal F}_{uu}+{\cal F}_{\phi\phi}=0 \;.
\ee
The simplest example of such an action is supplied by the free action
\be\label{4afree}
S_{free}=\frac{1}{2}\int dt d^4\theta\left( u^2 -\phi{}^2\right).
\ee

One may wonder whether the latter action is positively defined or not. After
passing to the component fields we find that, with the following
definition of the superspace integration measure
\be\label{4assint}
\int dt d^4\theta \equiv \frac{1}{24}\int dt D^{ia}D_{ja}D^{jb}D_{ib}\,,
\ee
and with the auxiliary fields eliminated by their equations of motion, the action \p{4afree}
yields the correct kinetic terms for the physical bosons
\be
S_{free} \sim \int dt \left( \frac{{\dot u}{}^2}{2} + \frac{{\dot \phi}{}^2}{2} +
\mbox{ fermions } \right).
\ee
The sign minus between two terms in \p{4afree} is actually related to the difference
between the defining constraints \p{4aconn4} for the superfields $u$ and $\phi$
and has no impact on the positive definiteness of the component action. Below we
shall meet more examples of such fake ``non-positive definiteness'' of the superfield
actions. The corresponding component actions are correct in all cases.

An interesting subclass of the actions \p{4aaction1} is provided
by those of the form:
\be\label{4aaction2}
S= \int dt d^4\theta \left[ f(u) +
\sum_{n=1}^\infty g_n(u) \phi^{2n} \right],
\ee
with the additional conditions (which follow from \p{4alaplace})
\be
2g_1=-f'',\quad g_{n+1}=-\frac{g''_n}{2(2n+1)(n+1)} \;,
\ee
where the primes  denote the derivatives with respect to $u$.
The immediate corollary is that {\it any action} written
in terms of the $N{=}4$ superfield $u$ can be promoted to an invariant
of $N{=}8$ supersymmetry by adding the appropriate interaction with the
superfield $\phi\,$.\footnote{One can equally start with an action of
the superfield $\phi$ and promote it to an $N{=}8$ supersymmetric action by turning on
the appropriate interactions with $u$.} For example, the action of $N{=}4$ SCQM \cite{leva2}
$$
S_{N{=}4}=\int dt d^4\theta\; u \log u
$$
can be generalized to have $N{=}8$ supersymmetry as
\bea
S_{N{=}8}&=&\int dt d^4\theta \left[ \frac{1}{2}u\log (u^2+\phi^2)
-\phi \arctan \left( \frac{\phi}{u} \right) \right] \nn
          &=&\frac{1}{2}  \int dt d^4\theta
\left[ \left( u+i\phi\right) \log  \left( u+i\phi\right) +
                  \left( u-i\phi\right) \log  \left( u-i\phi\right) \right].
\eea
Note that the $N{=}8$ superconformal invariance of this action is not
automatic and remains to be checked.

\vspace{0.5cm}

\noindent{\bf 2. } ({\bf 2, 8, 6}) = ({\bf 2, 4, 2}) $\oplus$ ({\bf 0, 4, 4})\\

There is a more sophisticated choice of a $N{=}4$ subspace in the
$N{=}8, d{=}1$ superspace, which gives rise to the second possible $N{=}4$
superfield splitting of the considered $N{=}8$ supermultiplet, that is
into the multiplets ({\bf 2, 4, 2}) and ({\bf 0, 4, 4}).

First of all, let us define a new set of covariant derivatives
\be\label{4bder}
\cD^{ia}=\frac{1}{\sqrt{2}}\left( D^{ia}-i\nabla^{ai}\right),\quad
\cbD^{ia}=\frac{1}{\sqrt{2}}\left( D^{ia}+i\nabla^{ai}\right),\quad
\left\{ \cD^{ia},\cbD^{jb}\right\}=2i\epsilon^{ij}\epsilon^{ab}\partial_t\,,
\ee
and new $N{=}8$ superfields $\cV, \cbV$ related to the original ones as
\be\label{4dsf}
\cV={\cal U}+i\Phi, \quad \cbV={\cal U}-i\Phi\;.
\ee
In this basis the constraints \p{31}, \p{32} read
\bea
&& \cD^{ia}\cV=0 ,\quad \cbD^{ia}\cbV=0, \nn
&& \cD^{i(a}\cD^{jb)}\cbV+\cbD^{i(a}\cbD^{j b)}\cV=0,\quad
\cD^{(i a}\cD^{j)b}\cbV-\cbD^{(i a}\cbD^{j)b}\cV=0. \label{4bcon1}
\eea
Now we split the complex quartet covariant derivatives \p{4bder}
into two sets of the doublet $N{=}4$ ones as
\be\label{4bder1}
D^i=\cD^{i1},\; \bD^{i}=\cbD^{i2},\quad \nabla^i=\cD^{i2},\;
\bar\nabla{}^i=-\cbD^{i1}
\ee
and cast the constraints \p{4bcon1} in the form
\bea
&& D^i\cV=0,\; \nabla^i\cV=0,\quad \bD_i\cbV=0,\; \overline\nabla_i\cbV=0 ,\nn
&& D^iD_i \cbV-{\overline\nabla}_i\overline\nabla{}^i\cV=0,
\quad D^i\nabla^j\cbV-\bD{}^i\overline\nabla{}^j\cV=0 \;. \label{4bcon2}
\eea
Next, as an alternative $N{=}4$ superspace, we choose the set of coordinates
closed under the action of $D^i, \bD{}^i$, i.e.
\be
\left( t\,,\; \theta_{i1} + i\vartheta_{i1}\,,\; \theta_{i2}
- i \vartheta_{i2} \right), \label{4bss}
\ee
while the $N{=}8$ superfields are expanded with respect to the
orthogonal combinations $\theta_{i1} - i \vartheta_{i1}$,
$\theta_{i2} + i\vartheta_{i2}$ which are annihilated by $D^i, \bD{}^i$.

As a consequence of the constraints \p{4bcon2}, the quadratic action
of the derivatives $\nabla^i$ and $\overline{\nabla}{}^i$ on every $N{=}8$
superfield $\cV, \cbV$  can be expressed as $D^i, \bD{}^i$
of some other superfield. Therefore, only
the zeroth and first order  components
of each $N{=}8$ superfield are independent
$N{=}4$ superfield projections. Thus, we are left with the following set
of $N{=}4$ superfields:
\be\label{4bn4sf}
\left. v=\cV\right|,\; \left. {\bar v}=\cbV\right|,\quad
\left. \psi^i=\overline\nabla{}^i\cV\right| ,\; \left.
\bpsi^i=-\nabla^i\cbV\right| \;.
\ee
These $N{=}4$ superfields prove to be subjected to the
additional constraints which also follow from \p{4bcon2}
\be\label{4bcon3}
D^i v=0\,,\;\; \bD{}^i {\bar v}=0\,,\quad D^i\psi^j=0\,,\;\;
\bD{}^i\bpsi{}^j=0\,,\;\; D^i\bpsi{}^j=-\bD{}^i\psi^j \;.
\ee
The $N{=}4$ superfields $v, {\bar v}$ comprise the standard
$N{=}4, d{=}1$ chiral multiplet ({\bf 2, 4, 2}),
while the $N{=}4$ superfields $\psi^i, \bpsi^j$ subjected to
\p{4bcon3} and both having the off-shell contents ({\bf 0, 4, 4})
are recognized as the fermionic version of the $N{=}4, d{=}1$
hypermultiplet.

The implicit $N{=}4$ supersymmetry is realized by the transformations
\bea\label{4btr}
&& \delta v=-\bar\eta{}^i \psi_i,\quad \delta\psi^i
=-\frac{1}{2}\bar\eta{}^i D^2 {\bar v}-2i\eta^i{\dot v}, \nn
&& \delta {\bar v}= \eta_i \bpsi^i, \quad \delta\bpsi_i
=\frac{1}{2}\eta_i \bD^2 v +2i\bar\eta_i{\dot {\bar v}}\:.
\eea

The invariant free action has the following form:
\be\label{4baction}
S_f=\int dt d^4\theta v{\bar v} -\frac{1}{2}\int dt d^2{\bar\theta}
\psi^i\psi_i-\frac{1}{2}\int dt d^2\theta \bpsi_i\bpsi{}^i\;.
\ee
Let us note that this very simple form of the action for the $N{=}4$ ({\bf 0, 4, 4})
supermultiplet $\psi_i,\bpsi{}^j$ is related to our choice of
the $N{=}4$ superspace. Indeed, with our definition \p{4bss}, the ({\bf 0, 4, 4})
supermultiplet is described by the doublet of chiral fermions with
the additional constraints \p{4bcon3} an essential part of which is simply the $N{=}4, d{=}1$
chirality conditions. Therefore, the chiral superspace is the best choice to write
the off-shell action of the multiplet ({\bf 0, 4, 4}) in the present case.
It is worthwhile to emphasize that all differently looking superspace off-shell actions of the
multiplet ({\bf 0, 4, 4}), viz. both terms in \p{1action}, the action \p{h5}
and the last two terms in \p{4baction}, yield the same component action for
this multiplet.

\subsection{Supermultiplet ({\bf 3, 8, 5})}
This supermultiplet has been discussed in detail in \cite{bikl} where
it was termed as the `$N{=}8$ tensor multiplet'. Here we shortly
remind its main features.

In the $N{=}8$ superspace this supermultiplet is described
by the triplet of bosonic superfields
$\cV^{ij}$ obeying the irreducibility constraints
\be\label{5con}
D_a^{(i}\cV^{jk)} =0 \; , \quad \nabla_\alpha{}^{(i}\cV^{jk)} =0 \; .
\ee
So three out of four original automorphism $SU(2)$ symmetries
remain manifest in this description.

The $N{=}8$ supermultiplet ({\bf 3, 8, 5}) can be decomposed into
$N{=}4$ supermultiplets in the two ways
\begin{itemize}
\item {\bf 1. } ({\bf 3, 8, 5}) = ({\bf 3, 4, 1}) $\oplus$ ({\bf 0, 4, 4})
\item {\bf 2. } ({\bf 3, 8, 5}) = ({\bf 1, 4, 3}) $\oplus$ ({\bf 2, 4, 2})
\end{itemize}

As in the previous case, we discuss both these options.\vspace{0.5cm}

\noindent{\bf 1. } ({\bf 3, 8, 5}) = ({\bf 3, 4, 1}) $\oplus$ ({\bf 0, 4, 4})\\

This  splitting requires choosing the coordinate set \p{ss1}
as the relevant $N{=}4$ superspace. Expanding the $N{=}8$ superfields $\cV^{ij}$
in $\vt_{i\alpha}$,  one finds that the constraints \p{5con} leave
in $\cV^{ij}$ the following four bosonic and four fermionic $N{=}4$ projections:
\be\label{5acomp}
\left. v^{ij}=\cV^{ij}\right| ,\quad \left. \xi^i_\alpha \equiv
\nabla_{j\alpha}\cV^{ij}\right|,\quad
\left. A\equiv \nabla^\alpha_i \nabla_{j\alpha}\cV^{ij}\right|
\ee
where $|$ means restriction to $\vt_{i\alpha}=0$.
As a consequence of \p{5con}, these $N{=}4$ superfields prove
to obey the constraints
\bea
&& D_a^{(i}v^{jk)}=0\,,\quad  D_a^{(i}\xi^{j)}_\alpha =0\,,\nn
&& A= 6m -D^a_i D_{aj} v^{ij}\,, \;m = \mbox{const}\, . \label{5acon1}
\eea
Thus, for the considered splitting, the $N{=}8$ tensor multiplet
superfield  $\cV^{ij}$ amounts to a direct sum of the $N{=}4$ `tensor' multiplet
superfield $v^{ij}$ with the  off-shell content $({\bf 3,\, 4,\, 1})$
and a fermionic $N{=}4$ hypermultiplet $\xi^i_\alpha$
with the off-shell content ({\bf 0, 4, 4}), plus a constant $m$
of the mass dimension.

The transformations of the implicit $N{=}4$ Poincar\'e supersymmetry
which together with the manifest $N{=}4$ supersymmetry constitute
the full off-shell $N{=}8$ Poincar\'e supersymmetry have the following
form in terms of the above $N{=}4$ superfields:
\be\label{5atransf1}
\delta^* v^{ij}=\eta^{(i}\xi^{j)}-\overline{\eta}{}^{(i}\bxi^{j)}\,,\quad
\delta^*\xi^i=-2i\overline{\eta}_j{\dot v}{}^{ij}
-\frac{1}{3}\overline{\eta}{}^i
D_j\bD_kv^{jk}+6m\,\overline{\eta}^i
\ee
where
\be
\eta^i \equiv \eta^{1i}\,, \;\; \overline{\eta}^i \equiv \eta^{2i}\,,\quad \label{complnot}
\xi^i \equiv \xi^{1i}\,,\;\; \bxi^i \equiv \xi^{2i} \,.
\ee
\vspace{0.5cm}

\noindent{\bf 2. } ({\bf 3, 8, 5}) = ({\bf 1, 4, 3}) $\oplus$ ({\bf 2, 4, 2})\\

This option corresponds to another choice of the $N{=}4$ superspace,
which amounts to dividing
the $N{=}8, d{=}1$ Grassmann coordinates into doublets with respect to some other
$SU(2)$ indices.
The relevant splitting of $N{=}8$ superspace into the $N{=}4$ subspace
and the complement of the latter can be performed  as follows. Firstly,
we define the new covariant derivatives  as
\bea\label{5bder}
&&  D^a\equiv \frac{1}{\sqrt{2}}\left( D^{1a}+i\nabla^{a1}\right),\;
\bD_a\equiv \frac{1}{\sqrt{2}}\left( D_a^{2}-i\nabla_a^{2}\right), \nn
&& \nabla^a\equiv \frac{i}{\sqrt{2}}\left( D^{2a}+i\nabla^{a2}\right),\;
\bar\nabla_a\equiv \frac{i}{\sqrt{2}}\left( D_a^{1}-i\nabla_a^{1}\right).
\eea
Then we choose the set of coordinates
closed under the action of $D^a, \bar D_a$, i.e.
\be
\left( t\,,\; \theta_{1a} - i\vartheta_{a1}\,,\;
\theta^{1a} + i \vartheta^{a1} \right), \label{ss2}
\ee
while the $N{=}8$ superfields are expanded with respect to
the orthogonal combinations $\theta^a_2 - i \vartheta^a_2\,$,
$\theta^a_1 + i\vartheta^a_1$ annihilated by $D^a, \bar D_a$.

The basic constraints \p{5con}, being rewritten in the basis \p{5bder},
take the form
\bea\label{5bcon1}
&& D^a\varphi=0\,, \quad D^a v -\nabla^a \varphi=0\,,
\quad \nabla^a v + D^a{\bar\varphi}=0\,,\quad \nabla^a \bar\varphi =0\,, \nn
&& {\overline{\nabla}}_a \varphi=0,\quad
\overline{\nabla}_a v+\bD_a \varphi=0\,,\quad
\bD_a v - \overline{\nabla}_a \bar\varphi =0\,,
\quad \bD_a\bar\varphi=0
\eea
where
\be\label{5bsf}
v\equiv -2i \cV^{12}\,,\quad \varphi \equiv \cV^{11}\,,\quad
\bar\varphi\equiv \cV^{22}\,.
\ee
Due to the constraints \p{5bcon1}, the derivatives $\nabla^a$
and $\overline{\nabla}_a$ of every $N{=}8$
superfield in the triplet $\left(\cV^{12}, \cV^{11}, \cV^{22} \right)$ can be
expressed as $D^a, \bD_a$ of some other superfield. Therefore, only
the zeroth order (i.e. taken at $\theta^a_2-i\vt_2^a=\theta^a_1+i\vt^a_1=0$)
components of each $N{=}8$ superfield are independent
$N{=}4$ superfield projections. These $N{=}4$ superfields are subjected
to the additional constraints which also follow from \p{5bcon1}
\be\label{5bcon2}
D^aD_a v=\bD_a \bD^a v=0\,,\quad D^a\varphi=0\,,\; \bD_a \bar\varphi=0\,.
\ee
The $N{=}4$ superfields $\varphi,\bar\varphi$ comprise the standard
$N{=}4, d{=}1$ chiral multiplet ({\bf 2, 4, 2}),
while the $N{=}4$ superfield $v$ subjected to \p{5bcon2} has the
needed off-shell content ({\bf 1, 4, 3}).

The implicit $N{=}4$ supersymmetry acts on the $N{=}4$ superfields
$v,\varphi,\bar\varphi$ as follows:
\be\label{5btransf3}
\delta^* v=\eta_a D^a \bar\varphi
+\bar\eta{}^a\bD_a \varphi\,,\quad \delta^*\varphi=-\eta_a D^a v\,,\quad
\delta^*\bar\varphi =-\bar\eta{}^a\bD_a v\,.
\ee

The invariant $N{=}4$ superfield actions for both decompositions of the
$N{=}8$ multiplet ({\bf 3, 8, 5}) were presented in Ref. \cite{bikl}.

Finally, we note that the considered multiplet can be treated as a dimensional
reduction of the $N{=}2, d{=}4$ tensor multiplet, whence its name `$N{=}8$ tensor multiplet'
used in \cite{bikl}. Three physical bosons of the $d{=}1$ case as compared to four bosonic
degrees of freedom of the $d{=}4$ case are related to the fact that one such degrees
in $d{=}4$ is represented by the `notoph'. After reduction to $d{=}1$ the notoph field strength
becomes a constant, and it is just the constant $m$ appearing in \p{5acon1}.

\subsection{Supermultiplet ({\bf 4, 8, 4})}

This supermultiplet can be described by a quartet of $N{=}8$ superfields
${\cal Q}^{a\alpha}$ obeying the constraints:
\be\label{6con}
D^{(a}_i{\cal Q}^{b)\alpha}=0,\quad \nabla^{(\alpha}_i{\cal Q}^{\beta)}_a=0 \;.
\ee
Let us note that the constraints \p{6con} are manifestly
covariant with respect to three $SU(2)$ subgroups realized
on the indices $i,a$ and $\alpha$.

{}From \p{6con} some important relations follow:
\be\label{6rel}
D^{ia}D^{jb}{\cal Q}^{c\alpha}
=2i\epsilon^{ij}\epsilon^{cb}{\dot{\cal Q}}{}^{a\alpha},\quad
\nabla^{i\alpha}\nabla^{j\beta}{\cal Q}^{a\gamma}
=2i\epsilon^{ij}\epsilon^{\gamma\beta}{\dot{\cal Q}}{}^{a\alpha}\;.
\ee
Using them, it is possible to show that the superfields
${\cal Q}^{a\alpha}$ contain the following independent components:
\be\label{6comp1}
\left. {\cal Q}^{a\alpha}\right|,\quad \left. D^i_a {\cal Q}^{a\alpha}\right|,
\quad \left.
\nabla^i_\alpha {\cal Q}^{a\alpha}\right|,\quad
\left. D^i_a \nabla^j_\alpha {\cal Q}^{a\alpha}\right|,
\ee
where $|$ means now restriction to $\theta_{ia}=\vt_{i\alpha}=0$.
This directly proves that we deal with the
irreducible ({\bf 4, 8, 4}) supermultiplet.

In order to construct the corresponding action, we pass to $N{=}4$ superfields.
There are three different possibilities to split this $N{=}8$ multiplet
into the $N{=}4$ ones:

\begin{itemize}
\item {\bf 1. } ({\bf 4, 8, 4}) = ({\bf 4, 4, 0}) $\oplus$ ({\bf 0, 4, 4})
\item {\bf 2. } ({\bf 4, 8, 4}) = ({\bf 3, 4, 1}) $\oplus$ ({\bf 1, 4, 3})
\item {\bf 3. } ({\bf 4, 8, 4}) = ({\bf 2, 4, 2}) $\oplus$ ({\bf 2, 4, 2})
\end{itemize}

Once again, we shall consider all these three cases separately.\vspace{0.5cm}

\noindent{\bf 1. } ({\bf 4, 8, 4}) = ({\bf 4, 4, 0}) $\oplus$ ({\bf 0, 4, 4})\\

This case implies the choice of the $N{=}4$ superspace \p{ss1}.
Expanding the $N{=}8$ superfields ${\cal Q}^{a\alpha}$
in $\vt_{i\alpha}$,  one may easily see that the constraints \p{6con}
leave in ${\cal Q}^{a\alpha}$ the following four bosonic and four
fermionic $N{=}4$ superfield projections:
\be\label{6acomp}
\left. q^{a\alpha}={\cal Q}^{a\alpha}\right| ,\quad
\left. \psi^{ia} \equiv \nabla^i_\alpha {\cal Q}^{a\alpha}\right|\,.
\ee
Each $N{=}4$ superfield is subjected, in virtue of \p{6con}, to an additional constraint
\be
 D^{i(a} q^{b)\alpha}=0\,,\quad  D^{i(a}\psi^{b)i} =0 . \label{6acon1}
\ee
Consulting Section~1, we come to the conclusion that these are just
the hypermultiplet $q^{i\alpha}$ with the off-shell field content ({\bf 4, 4, 0})
and a fermionic analog of the $N{=}4$ hypermultiplet $\psi^{ia}$ with the
field content ({\bf 0, 4, 4}).

The transformations of the implicit $N{=}4$ Poincar\'e supersymmetry
have the following form in terms of these $N{=}4$ superfields:
\be\label{6atransf1}
\delta^* q^{a\alpha}=\frac{1}{2}\eta^{i\alpha}\psi^a_i\,,\quad
\delta^* \psi^{ia}=-2i\eta^{i\alpha}{\dot q}{}^{a}_a\;.
\ee

Harmonic superspace provides the most adequate framework
for constructing the action for this splitting. We introduce
the harmonic variables parametrizing  the coset $SU(2)/U(1)$
as in eq. \p{h1}
and define the harmonic projections of $q^{a\alpha},\psi^{ai}$ by
\be
q^{+\alpha}=q^{a\alpha} u^+_a,\;  \psi^{+i}=\psi^{ia} u^+_a,\;.\label{hh2}
\ee
Now  the constraints \p{6acon1} take the standard form of the Grassmann
harmonic analyticity conditions
\be
D^{i+}q^{+\alpha}=0,\quad D^{i+}\psi^{+j}=0 \label{hh3}
\ee
where $D^{i+}=D^{ia}u^+_a$.
Thus, the superfields
$q^{\alpha +}$ and $ \psi^{i+}$ live in the analytic harmonic
$N{=}4, d{=}1$ superspace
$(\zeta, u^\pm_i) \equiv (t_A, \theta^{i+}, u^\pm_a)$.
In this setting, the transformations of the hidden $N{=}4$
supersymmetry \p{6atransf1} are rewritten as
\be\label{hh4}
\delta^* q^{\alpha +}=\frac{1}{2}\eta^{i\alpha}\psi^+_i,\quad
\delta^* \psi^{i+}=-2i\eta^{i\alpha}{\dot q}{}^+_\alpha \;.
\ee
The free action is
\be\label{hh5}
S=\frac{i}{2}\int du d\zeta^{--}
\left( q^{\alpha +}{\dot q}{}^+_\alpha -\frac{i}{4}
\psi^{i+}\psi^+_i \right).
\ee
where $du d\zeta^{--}= du dt_Ad\theta^{i+}d\theta^+_i $ is the measure
of integration over the analytic superspace.\vspace{0.5cm}

\noindent{\bf 2. } ({\bf 4, 8, 4}) = ({\bf 3, 4, 1}) $\oplus$ ({\bf 1, 4, 3})\\

In order to describe this $N{=}4$ superfield realization of the
$N{=}8$ supermultiplet $({\bf 4,\, 8,\, 4})$, we introduce the $N{=}8$
superfields ${\cal V}{}^{ab}, {\cal V}$ as
\be\label{6csf}
{\cal Q}{}^{a\alpha} \equiv \delta^\alpha_b{\cal V}{}^{a b}
-\epsilon^{a\alpha}{\cal V}\,, \quad {\cal V}{}^{a b} = {\cal V}{}^{b a}\,,
\ee
and use the covariant derivatives \p{4bder} to rewrite
the basic constraints \p{6con} as
\bea
&& \cD^{i(a}{\cal V}^{bc)}=0,
\quad \cbD{}^{i(a}{\cal V}^{bc)}=0, \label{6ccon1} \\
&& \cD^{ia}{\cal V}=\frac{1}{2}\cbD{}^i_b{\cal V}^{ab},\quad
\cbD{}^{ia}{\cal V}=\frac{1}{2}\cD^i_b{\cal V}^{ab}\;. \label{6ccon2}
\eea
The constraints \p{6ccon1} define ${\cal V}^{ab}$ as the $N{=}8$
superfield comprising the off-shell multiplet ({\bf 3, 8, 5}),
while, as one can deduce from \p{6ccon1}, \p{6ccon2}, the $N{=}8$
superfield ${\cal V}$ has the content ({\bf 1, 8, 7}). Then the constraints
\p{6ccon2} establish relations between the fermions in these two superfields
and reduce the number of independent auxiliary fields to four, so that
we once again end up with the irreducible $N{=}8$ multiplet ({\bf 4, 8, 4}).

Two sets of $N{=}4$ covariant derivatives
$$\left( D^a,\bD{}^a\right) \equiv
\left( \cD^{1a},\cbD^{2a}\right) \mbox{ and  }
\left( \bar\nabla{}^a,\nabla^a\right) \equiv \left( \cD^{2a},\cbD^{1a}\right)$$
 are naturally realized in terms of the
$N{=}4$ superspaces
$\left( t,\theta_{1a}+i\vt_{1a},\theta_{2a}-i\vt_{2a}\right)$ and
$\left( t, \theta_{2a}+i\vt_{2a}\right. ,$ $\left.
\theta_{1a}-i\vt_{1a}\right)$.
In terms of the new derivatives the constraints \p{6ccon1}, \p{6ccon2}
become
\bea
&& D^{(a}\cV^{bc)}=\bD{}^{(a}\cV^{bc)}
=\nabla^{(a}\cV^{bc)}=\bar\nabla{}^{(a}\cV^{bc)}=0 , \nn
&& D^a \cV=\frac{1}{2}\nabla_b \cV^{ab},\; \bD{}^a \cV
=\frac{1}{2}\bar\nabla_b \cV^{ab},\quad
\nabla^a \cV=\frac{1}{2}D_b \cV^{ab},\;
\bar\nabla{}^a \cV=\frac{1}{2}\bD_b \cV^{ab}.\label{6ccon3}
\eea
Now we see that the $\nabla^a, \bar\nabla_a$ derivatives of the
superfields $\cV, \cV^{ab}$ are expressed as $D^a,\bD{}^a$
of the superfields $\cV^{ab},\cV\,$, respectively. Thus, in the
$\left( \theta_{2a}+i\vt_{2a},\theta_{1a}-i\vt_{1a}\right)$
expansions of the superfields
$\cV,\cV^{ab}$ only the first components (i.e. those of zero order in
the coordinates $\left(\theta_{2a}+i\vt_{2a},
\theta_{1a}-i\vt_{1a}\right)$) will be independent $N{=}4$ superfields.
We denote them  $v, v^{ab}$. The hidden $N{=}4$ supersymmetry is realized
on these $N{=}4$ superfields as:
\be\label{6ctr1}
\delta v =-\frac{1}{2} \eta_a D_b v^{ab}+\frac{1}{2}\bar\eta_a\bD_b v^{ab},\;
\delta v^{ab}=\frac{4}{3}\left( \eta^{(a}D^{b)}v
-\bar\eta{}^{(a}\bD{}^{b)}v\right),
\ee
while the superfields themselves obey the constraints
\be\label{6ccon4}
D^{(a}v^{bc)}=\bD{}^{(a}v^{bc)}=0,\quad D^{(a}\bD{}^{b)}v=0\;,
\ee
which are remnant of the $N{=}8$ superfield constraints \p{6ccon3}.

The invariant free action reads
\be\label{6caction}
S=\int dt d^4\theta \left( v^2 -\frac{3}{8} v^{ab}v_{ab}\right).
\ee
\vspace{0.5cm}

\noindent{\bf 3. } ({\bf 4, 8, 4}) = ({\bf 2, 4, 2}) $\oplus$ ({\bf 2, 4, 2})\\

This case is a little bit more tricky. First of all we define the
new set of $N{=}8$ superfields $\cW,\Phi$ in terms of $\cV^{ij},\cV$
defined earlier in \p{6csf}
\be\label{6bsf}
\cW \equiv \cV^{11}, \overline\cW\equiv \cV^{22},
\quad \Phi\equiv\frac{2}{3}\left( \cV+\frac{3}{2}\cV^{12}\right),\;
\overline\Phi\equiv\frac{2}{3}\left( \cV-\frac{3}{2}\cV^{12}\right)
\ee
and construct two new sets of $N{=}4$ derivatives $D^i,\nabla^i$
{}from those defined in \p{4bder}:
\bea\label{6bder}
&&D^i=\frac{1}{\sqrt{2}}\left(\cD^{i1}+\cbD{}^{i1}\right),\;
\bD{}^i=\frac{1}{\sqrt{2}}\left(\cD^{i2}+\cbD{}^{i2}\right),\nn
&& \nabla^i=\frac{1}{\sqrt{2}}\left(\cD^{i1}-\cbD{}^{i1}\right),\;
\overline\nabla{}^i=-\frac{1}{\sqrt{2}}\left(\cD^{i2} - \cbD{}^{i2}\right).\;
\eea

The basic constraints \p{6ccon1}, \p{6ccon2} can be rewritten in terms
of the superfields $\cW,\Phi$ and the derivatives $D^i, \nabla^i$ as
\bea\label{6bcon1}
&& D^i \cW= \nabla^i \cW=0,\; \bD{}^i \overline\cW
=\overline\nabla{}^i \overline\cW=0,\quad   D^i \Phi
=\overline\nabla{}^i \Phi =0 ,\;
 \nabla^i \overline\Phi=\bD{}^i\overline\Phi=0,\nn
&& \bD{}^i \cW - D^i \overline\Phi=0,\;
D^i \overline\cW+\bD{}^i \Phi=0,\quad
\overline\nabla{}^i \cW-\nabla^i \Phi=0,\;
\nabla^i \overline\cW+\overline\nabla{}^i \overline\Phi=0.
\eea
The proper $N{=}4$ superspace is defined as the one on
which the covariant derivatives $D^1, \bD{}^2, \nabla^1,
\overline\nabla{}^2$
are naturally realized. The constraints \p{6bcon1} imply
that the remaining set of covariant derivatives, i.e.
$D^2, \bD{}^1, \nabla^2, \overline\nabla{}^1$,
when acting on every involved $N{=}8$ superfield, can be expressed
as spinor derivatives from the the first set acting on some
another $N{=}8$ superfield. Thus the first $N{=}4$
superfield components of the
$N{=}8$ superfields $\cW, \Phi$ are the only independent $N{=}4$
superfield projections. The transformations of the implicit
$N{=}4$ Poincar\'e supersymmetry have the following form in terms of
these $N{=}4$ superfields:
\bea\label{6btransf}
&& \delta w = \beps D^1 \bar\phi +
\bar\eta \nabla^1 \phi,\quad \delta \phi
= -\eta\overline\nabla_1 w -\beps D^1 {\bar w}\;, \nn
&& \delta {\bar w} = \epsilon \bD_1 \phi
+ \eta \overline\nabla_1{\bar\phi},\quad
\delta{\bar\phi}=-\epsilon\bD_1 w-\bar\eta \nabla^1 {\bar w}\;.
\eea

The free invariant action is
\be\label{6baction}
S=\int dt d^4 \theta \left( w{\bar w}-\phi\bar\phi \right).
\ee

\subsection{Supermultiplet ({\bf 5, 8, 3})}
This supermultiplet has been considered in details
in Refs. \cite{{DE},{bikl}}. It was termed there the `$N{=}8$ vector multiplet'.
Here we sketch its main properties.

In order to describe this supermultiplet, one should introduce
five bosonic $N{=}8$ superfields ${\cal V}_{\alpha a},{\cal U}$
obeying the constraints
\be\label{5constr}
D^{ib}{\cal V}_{\alpha a} + \delta_a^b \nabla_\alpha^i{\cal U}=0\;, \quad
\nabla^{\beta i}{\cal V}_{\alpha a} + \delta_\alpha^\beta D^i_a {\cal U}=0
\;.
\ee

It is worth noting that the constraints \p{5constr} are covariant
not only under three $SU(2)$ automorphism groups (realized on the
doublet indices $i$, $a$ and $\alpha$), but also under the $SO(5)$
automorphisms. These $SO(5)$ transformations mix the spinor derivatives
$D^{ia}$ and $\nabla^{\alpha i}$ in the indices $\alpha$ and $a$,
while two $SU(2)$ groups realized on these indices constitute $SO(4)\subset SO(5)$.
The superfields $\cU,\cV^{\alpha a}$ form an $SO(5)$ vector:
under the $SO(5)$ transformations belonging to the coset $SO(5)/SO(4)$
they transform  as
\be\label{so5a}
\delta \cV_{\alpha a}= a_{\alpha a}\; \cU \;,
\quad \delta \cU=-2 a_{\alpha a}\; \cV^{\alpha a}\,.
\ee

As in the previous cases we may consider two different splittings
of the $N{=}8$ vector multiplet into irreducible $N{=}4$ superfields
\begin{itemize}
\item {\bf 1. } ({\bf 5, 8, 3}) = ({\bf 1, 4, 3}) $\oplus$ ({\bf 4, 4, 0})
\item {\bf 2. } ({\bf 5, 8, 3}) = ({\bf 3, 4, 1}) $\oplus$ ({\bf 2, 4, 2})
\end{itemize}
Once again, they correspond to two different choices of
the $N{=}4, d{=}1$ superspace as a subspace in the original $N{=}8, d{=}1$ superspace.
\vspace{0.5cm}

\noindent{\bf 1. } ({\bf 5, 8, 3}) = ({\bf 1, 4, 3}) $\oplus$ ({\bf 4, 4, 0})\\

The relevant $N{=}4$ superspace is $\mathbb{R}^{(1|4)}$ parametrized by the coordinates
$\left( t, \theta_{ia} \right)$ and defined in \p{ss1}. As in the previous cases,
it follows from the constraints \p{5constr} that the spinor derivatives of
all involved superfields with respect to $\vartheta_{i\alpha}$
are expressed in terms of spinor derivatives with respect to
$\theta_{i a}$. Thus the only essential $N{=}4$ superfield components
of $\cV_{\alpha a}$ and $\cU$ in their $\vartheta$-expansion are the
first ones
\be\label{n4comp}
v_{\alpha a} \equiv \cV_{\alpha a}|_{\vt=0}\;,\quad u \equiv \cU|_{\vt=0}\,.
\ee
They accommodate the whole off-shell component content of
the $N{=}8$ vector multiplet. These five bosonic $N{=}4$ superfields are
subjected, in virtue of \p{5constr}, to the irreducibility constraints
in $N{=}4$ superspace
\be\label{5constra}
D^{i(a} v^{b)\alpha}=0,\quad D^{i(a}D_i^{b)} u= 0.
\ee
Thus, from the $N{=}4$ superspace standpoint, the vector $N{=}8$ supermultiplet
is the sum of the $N{=}4,d{=}1$ hypermultiplet $ v_{\alpha a}$ with
the  off-shell component contents ({\bf 4, 4, 0}) and the
$N{=}4$ `old' tensor multiplet $u$ with the contents ({\bf 1, 4, 3}).

The transformations of the implicit $N{=}4$ Poincar\'e supersymmetry are
\be\label{5n4transfa}
\delta v_{a\alpha}= \eta_{i\alpha} D^i_a u\;, \quad \delta u
={1\over 2}\eta_{i\alpha}D^{ia} v_a^\alpha\;.
\ee
\vspace{0.5cm}

\noindent{\bf 2. } ({\bf 5, 8, 3}) = ({\bf 3, 4, 1}) $\oplus$ ({\bf 2, 4, 1})\\

Another interesting $N{=}4$ superfield splitting of the $N{=}8$
vector multiplet can be achieved by passing to the complex parametrization
of the $N{=}8$ superspace as
$$
\left(t,\Theta_{i\alpha}= \theta_{i\alpha}+i\vartheta_{\alpha i},
\bar\Theta^{i\alpha}= \theta^{i\alpha} -i\vartheta^{\alpha i}\right)
$$
where we have identified the indices $a$ and $\alpha$, thus having reduced the number of
manifest $SU(2)$ automorphism symmetries to just two. In this superspace
the covariant derivatives $\cD^{i\alpha},\cbD{}^{j\beta}$
defined in \p{4bder} (with the identification of indices just mentioned)
are naturally realized. We are also led to define new superfields
\bea
\cV \equiv -\epsilon_{\alpha a}\cV^{\alpha a}\,,
\quad \cW^{\alpha\beta} \equiv \cV^{(\alpha \beta)} = {1\over 2}
\left( \cV^{\alpha \beta}+\cV^{\beta \alpha}\right)\,,
\quad \cW \equiv \cV +i{\cal U}\,,
\quad \overline{\cW} \equiv \cV - i{\cal U}\,.
\eea

In this basis of $N{=}8$ superspace the original constraints
\p{5constr} amount to
\bea\label{5constrb}
&&\cD^{i\alpha}\cW^{\beta\gamma}=
-\frac{1}{4}\left( \epsilon^{\beta\alpha}\cbD{}^{i\gamma}\overline{\cW}+
\epsilon^{\gamma\alpha}\cbD{}^{i\beta}\overline{\cW}\right),\;
\cbD{}^{i\alpha}\cW^{\beta\gamma}=-\frac{1}{4}\left( \epsilon^{\beta\alpha}
\cD^{i\gamma} \cW+\epsilon^{\gamma\alpha}\cD{}^{i\beta}\cW\right),\nn
&& \cD^{i\alpha}\overline{\cW}=0,\;\cbD^{i\alpha} \cW=0\,,
\quad (\cD^{k\alpha}\cD^i_\alpha) \cW
=(\cbD^{k}_{\alpha}\cbD^{i\alpha})\overline{\cW}\,.
\eea
Next, we single out the $N{=}4, d{=}1$ superspace  as
$\left( t, \theta_\alpha \equiv \Theta_{1\alpha}, \bar\theta^{\alpha}\right)$
and split our $N{=}8$ superfields into the $N{=}4$ ones in the standard way.
As in all previous cases,  the spinor derivatives of
each $N{=}8$ superfield with respect to
$\overline{\Theta}^{2\alpha}$ and $\Theta_{2\alpha}$, as
a consequence of the constraints \p{5constrb}, are expressed as derivatives
of some other superfields with respect to $\bar\theta^{\alpha}$
and $\theta_{\alpha}$. Therefore, only the first
(i.e. taken at $\overline{\Theta}^{2a}=0$ and $\Theta_{2a}=0$)
$N{=}4$ superfield components of the $N{=}8$ superfields really matter.
They accommodate the entire off-shell field content of the multiplet.
These $N{=}4$ superfields are defined as
\be
\left. \phi \equiv \cW \right|\,,
\quad, \left. \bar\phi \equiv \overline{\cW}\right|\,,
\quad \left. w^{\alpha\beta} \equiv \cW^{\alpha\beta}\right|
\ee
and satisfy the constraints following from \p{5constrb}
\be\label{5finalconstr}
{\cD}^{\alpha} \bar\phi =0\,,\quad \cbD_{\alpha}\phi=0\,,\quad
\cD^{(\alpha}w^{\beta\gamma)}=\cbD{}^{(\alpha}w^{\beta\gamma)}=0\,,
\quad \cD^\alpha
\equiv \cD^{1\alpha}\,, \;\cbD_\alpha \equiv \cbD_{1\alpha}\,.
\ee
They tell us that the $N{=}4$ superfields $\phi$ and $\bar\phi$
form the standard $N{=}4$ chiral multiplet ({\bf 2, 4, 2}),
while the $N{=}4$ superfield $w^{\alpha\beta}$ represents the $N{=}4$
tensor multiplet ({\bf 3, 4, 1}).

The implicit $N{=}4$ supersymmetry is realized on $w^{\alpha\beta}$,
$\phi$ and $\bar\phi$ as
\be\label{6tra}
\delta w^{\alpha\beta}=\frac{1}{2}\left(\eta^{(\alpha}\cbD^{\beta)}\bar\phi
-\bar\eta{}^{(\alpha}\cD{}^{\beta)}\phi\right)\,,\quad \delta \phi=
\frac{4}{3}\eta_\alpha \cbD^\beta w_\beta^\alpha\,,\quad
\delta  \bar\phi=-\frac{4}{3}\bar\eta{}^{\alpha}\cD_\beta w_\alpha^\beta\,.
\ee

An analysis of $N{=}8$ supersymmetric actions for the $N{=}8$ vector multiplet
may be found in \cite{bikl}. Here we present only two examples of such actions.

The first action
\be\label{dok5a}
S= 2\int dt d^4\theta\,\frac{1}{{\hat u} +\sqrt{{\hat{u}{}^2}+2\hat{v}{}^2}}
\ee
corresponds to the splitting {\bf 1} above. It
possesses both $N{=}4$ superconformal symmetry and $N{=}8$ Poincar\'e
supersymmetry. The closure of these two supersymmetries yields the whole
$N{=}8$ conformal supersymmetry associated with
the supergroup $OSp(4^\star |4)$ \cite{bikl}.

The second action is constructed using the splitting {\bf 2}.
It  includes the $N{=}4$ tensor $w^{\alpha\beta}$ and the chiral
$\phi$, $\bar\phi$ multiplets. The free action invariant under
the implicit $N{=}4$ supersymmetry \p{6tra} reads
\be\label{6actiona1}
S_{free}=\int dt d^4\theta \left( w^2-\frac{3}{4}\phi\bar\phi\right).
\ee
The $N{=}4$ superconformally invariant action which is
also invariant under \p{6tra} has the very simple form
\be\label{6actiona2}
S_{kin}=2\int dt d^4 \theta\;{{\log \left( \sqrt{w^2}+
\sqrt{w^2+\frac{1}{2}\phi\bar\phi}\right)}\over\sqrt{w^2}}\,.
\ee
It is also $OSp(4^\star \vert 4)$ invariant.

Finally, we note that the multiplet ({\bf 5, 8, 3}) can be regarded as a
dimensional reduction of the abelian $N{=}2, d{=}4$ gauge multiplet \cite{DE}
(this is the reason why sometimes it is referred to as `$N{=}8$ vector multiplet' \cite{bikl}).
Three extra physical scalar fields of the $d{=}1$ case come from the spatial components of the
$d{=}4$ gauge vector potential which become gauge invariant after reduction to $d{=}1$.
The description of this multiplet in $N{=}8, d{=}1$ harmonic superspace was given in
\cite{{BMZ},{ISmi}}.

\subsection{Supermultiplet ({\bf 6, 8, 2})}
This supermultiplet can be described by two $N{=}8$ tensor
multiplets $\cV^{ij}$ and $\cW^{ab}\,$,
\be\label{7con1}
D^{(i}_a\cV^{jk)}=0\,,\; \nabla^{(i}_a\cV^{jk)}=0\,,\quad
D^{(a}_i\cW^{bc)}=0\,,\; \nabla^{(a}_i\cW^{bc)}=0\,,
\ee
with the additional constraints
\be\label{7con2}
D^a_j \cV^{ij}=\nabla^{bi}\cW^a_b,\quad \nabla^a_j \cV^{ij}=-D^i_b \cW^{ab} \;.
\ee
The role of the latter constraints is to identify the eight fermions
which are present in $\cV^{ij}$ with the fermions from
$\cW^{ab}$, and to reduce the number of independent auxiliary fields
in both superfields to two
\be\label{7auxil}
F_1=D^a_iD_{aj}\cV^{ij}|\,,\quad F_2 = D^i_a D_{ib}\cW^{ab}|\,,
\ee
where $|$ means here restriction to $\theta_{ia}=\vt_{ia}=0\,$.

There are two different possibilities to split this $N{=}8$ multiplet
into the $N{=}4$ ones

\begin{itemize}
\item {\bf 1. } ({\bf 6, 8, 2}) = ({\bf 3, 4, 1}) $\oplus$ ({\bf 3, 4, 1})
\item {\bf 2. } ({\bf 6, 8, 2}) = ({\bf 4, 4, 0}) $\oplus$ ({\bf 2, 4, 2})
\end{itemize}

As before, we discuss peculiarities of both decompositions.
\vspace{0.5cm}

\noindent{\bf 1. } ({\bf 6, 8, 2}) = ({\bf 3, 4, 1}) $\oplus$ ({\bf 3, 4, 1})\\

The corresponding $N{=}4$ supersubspace is \p{ss1}.
The $N{=}8$ constraints imply that the only
essential $N{=}4$ superfield components
of $\cV^{ij}$ and $\cW^{ab}$ in their $\vartheta$-expansion are the first ones
\be\label{7comp1}
v^{ij} \equiv \cV^{ij}|\;,\quad w^{ab} \equiv \cW^{ab}| \,.
\ee
These six bosonic $N{=}4$ superfields are
subjected, in virtue of \p{7con1}, \p{7con2}, to the irreducibility
constraints in $N{=}4$ superspace
\be\label{7con3}
D^{a(i} v^{jk)}=0\,,\quad D^{i(a} w^{bc)}= 0\,.
\ee
Thus, the  $N{=}8$ supermultiplet ({\bf 6, 8, 2}) amounts to the sum of
two $N{=}4,d{=}1$ tensor multiplets $v^{ij}, w^{ab}$ with
the off-shell field contents ({\bf 3, 4, 1}) $\oplus$ ({\bf 3, 4, 1}).

The transformations of the implicit $N{=}4$ Poincar\'e supersymmetry
are
\be\label{7transf1}
\delta v^{ij}=-\frac{2}{3}\eta^{(i}_a D^{j)}_b w^{ab},\quad
\delta w^{ab}=\frac{2}{3} \eta^{(a}_i D^{b)}_j v^{ij}.
\ee

The free $N{=}8$ supersymmetric action has the following form:
\be\label{7action1}
S=\int dt d^4 \theta \left( v^2- w^2 \right).
\ee
\vspace{0.5cm}

\noindent{\bf 2. } ({\bf 6, 8, 2}) = ({\bf 4, 4, 0}) $\oplus$ ({\bf 2, 4, 2})\\

In this case, to describe the ({\bf 6, 8, 2}) multiplet,
we combine two $N{=}4$ superfields: chiral
superfield
\be\label{7cona}
D^i \phi=\bD{}^i \bar\phi=0
\ee
and the hypermultiplet $q^{ia}$
\be\label{7cona1}
D^{(i} q^{j)a}=\bD{}^{(i} q^{j)a}=0 \:.
\ee

The transformations of implicit $N{=}4$ supersymmetry read
\be\label{7tra}
\delta q^{ia}=\beps{}^a D^i\bar\phi +\epsilon^a \bD{}^i \phi\;, \quad
\delta\phi=-\frac{1}{2}\beps{}^a D^i q_{ia}\;, \;
\delta\bar\phi=-\frac{1}{2}\epsilon^a\bD{}^i q_{ia}\;.
\ee

The invariant free action is as follows:
\be\label{7actiona1}
S_{free}=\int dt d^4\theta \left( q^2-4\phi\bar\phi\right)\;.
\ee

\subsection{Supermultiplet ({\bf 7, 8, 1})}
This supermultiplet has a natural description in terms of
two $N{=}8$ superfields $\cV^{ij}$ and $\cQ^{a\alpha}$ satisfying
the constraints
\bea
&& D^{(i a}\cV^{jk)}=0,\quad \nabla^{\alpha(i}\cV^{jk)}=0, \quad
 D^{i(a}\cQ^{\alpha b)}=0, \quad \nabla^{(\alpha}_i\cQ^{\beta)}_a=0,
\label{8con1} \\
&& D_j^a \cV^{ij}=i\nabla^i_\alpha \cQ^{a\alpha},
\quad \nabla^\alpha_j \cV^{ij}=-iD^i_a \cQ^{a\alpha} .\label{8con2}
\eea
The constraints \p{8con1} leave in the superfields $\cV^{ij}$
and $\cQ^{a\alpha}$ the sets ({\bf 3, 8, 5}) and $({\bf 4,\, 8,\, 4})$
of irreducible components, respectively. The role of the
constraints \p{8con2} is to identify the fermions in the
superfields $\cV^{ij}$ and $\cQ^{a\alpha}$ and to reduce the
total number of independent auxiliary components in both
superfields to just one.

For this supermultiplet there is a unique splitting into
$N{=}4$ superfields as
$$
({\bf 7,\, 8,\, 1}) = ({\bf 3,\, 4,\, 1}) \oplus ({\bf 4,\, 4,\, 0}).
$$
The proper $N{=}4$ superspace is parametrized by the coordinates
$\left( t, \theta_{ia} \right)$.
The constraints \p{8con1}, \p{8con2} imply that the only
essential $N{=}4$ superfield components in the  $\vartheta$-expansion
of $\cV^{ij}$ and $\cQ^{a\alpha}$ are the first ones
\be\label{8comp}
v^{ij} \equiv \cV^{ij}|_{\vt=0}\;,\quad
q^{a\alpha} \equiv \cQ^{a\alpha}|_{\vt=0}\,.
\ee
These seven bosonic $N{=}4$ superfields are subjected, as a corollary of
\p{8con1}, \p{8con2}, to the irreducibility constraints in $N{=}4$ superspace
\be\label{8const}
D^{a(i} v^{jk)}=0\,,\quad D^{i(a} q^{b)\alpha} = 0\,.
\ee
Thus the $N{=}8$ supermultiplet ({\bf 7, 8, 1}) amounts to the sum of the
$N{=}4, d{=}1$ hypermultiplet $ q^{a\alpha }$ with
the ({\bf 4, 4, 0}) off-shell field content and the
$N{=}4$ tensor multiplet $v^{ij}$ with the ({\bf 3, 4, 1}) content.

The implicit $N{=}4$ Poincar\'e supersymmetry is realized by the
transformations
\be\label{8transfa}
\delta v^{ij}= -\frac{2i}{3}\,\eta^{(i}_{\alpha} D^{j)}_a q^{a\alpha}\;, \quad
\delta q^{a\alpha}
=-{i\over 2}\eta^{i\alpha}D^{ja} v_{ij}\;.
\ee

The free action can be also easily written
\be\label{8action}
S=\int dt d^4\theta \left[ v^2 -\frac{4}{3} q^2  \right].
\ee

\subsection{Supermultiplet ({\bf 8, 8, 0})}

This supermultiplet is analogous to the supermultiplet ({\bf 0, 8, 8}):
they differ in their overall Grassmann parity. It is described by the two
real bosonic $N{=}8$ superfields
$\cQ^{aA}, \Phi^{i\alpha}$ subjected to the constraints
\bea
&& D^{(i a}\Phi^{j)\alpha}=0\,,\; D^{i(a} \cQ^{b)A}=0\,,\quad
\nabla^{(\alpha A}\Phi^{\beta)}_i=0\,,\; \nabla^{\alpha( A} \cQ^{a B)}=0\,,
\label{9con1} \\
&& \nabla^{\alpha A} \cQ_A^a=-D^{ia}\Phi_i^\alpha\,,\quad
\nabla^{\alpha A}\Phi_\alpha^i=D^{ia} \cQ_a^A\,. \label{9con2}
\eea

Analogously to the case of the supermultiplet ({\bf 8, 8, 0}),
{}from the constraints \p{9con2} it follows that the spinor derivatives of
all involved superfields with respect to $\vartheta_{\alpha A}$
are expressed in terms of spinor derivatives with respect to
$\theta_{i a}$. Thus the only essential $N{=}4$ superfield components
in the $\vartheta$-expansion
of $\cQ^{aA}$ and $\Phi^{i\alpha}$ are the first ones
\be\label{9comp1}
q^{aA} \equiv \cQ^{aA}|_{\vt=0}\;,
\quad \phi^{i\alpha} \equiv \Phi^{i\alpha}|_{\vt=0}\,.
\ee
They accommodate the whole off-shell component content of the
multiplet ({\bf 8, 8, 0}). These  bosonic $N{=}4$ superfields are
subjected, as a consequence of \p{9con1}, \p{9con2},
to the irreducibility constraints in $N{=}4$ superspace
\be\label{9con3}
D^{a(i}\phi^{j)\alpha}=0\,,\; D^{i(a}q^{b)A}=0\,.
\ee
Thus the  $N{=}8$ supermultiplet ({\bf 8, 8, 0}) can be represented
as the sum of two $N{=}4, d{=}1$  hypermultiplets  with
the off-shell component contents ({\bf 4, 4, 0}) $\oplus$ ({\bf 4, 4, 0}).

The transformations of the implicit $N{=}4$ Poincar\'e supersymmetry
in this last case are as follows:
\be\label{9tr1}
\delta q^{aA}=-\frac{1}{2}\eta^{A\alpha} D^{ia} \phi_{i\alpha},\quad
\delta \phi^{i\alpha}=\frac{1}{2}\eta^\alpha_A D^i_a q^{aA}.
\ee

The invariant free action is
\be\label{9action1}
S=\int dt d^4\theta \left[ q^2 - \phi^2\right].
\ee

The most general action still respecting four $SU(2)$ automorphism symmetries
has the following form:
\be\label{9action2}
S=\int dt d^4\theta F( q^2, \phi^2),
\ee
where, as the necessary condition of $N{=}8$ supersymmetry,
the function $F( q^2, \phi^2)$ should obey the equation
\be\label{9action2a}
\frac{\partial^2}{\partial q^2 \partial q^2} \left( q^2 F( q^2, \phi^2)\right)+
\frac{\partial^2}{\partial \phi^2 \partial \phi^2}
\left( \phi^2 F( q^2,\phi^2)\right) = 0.
\ee

\section{Summary and conclusions}
In this paper, as a further development of findings of our previous
paper \cite{bikl}, we presented superfield formulations
of the full amount of off-shell $N{=}8, d{=}1$ supermultiplets
with {\bf 8} physical fermions, both in $N{=}8$ and $N{=}4$ superspaces.
We listed all possible $N{=}4$ superfield splittings of these multiplets.
For each such splitting we gave the $N{=}8$ supersymmetric free $N{=}4$
superfield action, and for some special cases we quoted
examples of $N{=}8$ supersymmetric actions with interaction. Only two of
these $d{=}1$ multiplets, ({\bf 3, 5, 8}) and ({\bf 5, 8, 3}) considered
earlier in \cite{bikl}, can be understood in the framework of the dimensional
reduction $N{=}2,\, d{=}4 \rightarrow N{=}8,\, d{=}1$. It is far from obvious
whether the $N{=}8,\, d{=}1$ off-shell superfield formulations of other multiplets considered
here can be recovered in a similar way.\footnote{Perhaps one should start
{}from an enlarged set of $N{=}4, d{=}4$ off-shell superfields, with broken
Lorentz covariance.} Fortunately, there is no actual need to care about this since
our approach is self-consistent in $d{=}1$.

Our results are summarized in the Table below.

\begin{center}
{\bf N=8 supermultiplets }
\vspace{0.4cm}

\begin{tabular}{|c|c|c|l|}
\hline
Multiplet & $N{=}8$ Superfields & $N{=}8$ Constraints & $N{=}4$ splittings  \\
\hline
({\bf 0, 8, 8}) & $\Psi{}^{aA},\Xi{}^{i\alpha}$  &  \p{1con1}, \p{1con2} & $\begin{array}{l}
                     ({\bf 0,\,4,\,4}) \oplus ({\bf 0,\,4,\,4}) \end{array}$\\ \hline
({\bf 1, 8, 7}) & $\cU$  &  \p{2con1a}, \p{2con1b} & $\begin{array}{l} ({\bf 1,\,4,\,3}) \oplus ({\bf 0,\,4,\,4})
                                        \end{array}$\\ \hline
({\bf 2, 8, 6}) & $\cU, \Phi$ &  \p{31}, \p{32} & $ \begin{array}{l}   ({\bf 1,\,4,\,3}) \oplus ({\bf 1,\,4,\,3})\\
                                                                  ({\bf 2,\,4,\,2}) \oplus ({\bf 0,\,4,\,4})
                                                \end{array} $\\ \hline
({\bf 3, 8, 5}) & $\cV{}^{ij}$ &  \p{5con}  &  $ \begin{array}{l}   ({\bf 3,\,4,\,1}) \oplus ({\bf 0,\,4,\,4})\\
                                                                  ({\bf 1,\,4,\,3}) \oplus ({\bf 2,\,4,\,2})
                                                \end{array} $\\\hline
({\bf 4, 8, 4}) & $\cQ{}^{a\alpha}$ &  \p{6con} &  $\begin{array}{l}   ({\bf 4,\,4,\,0}) \oplus ({\bf 0,\,4,\,4})\\
                                          ({\bf 3,\,4,\,1}) \oplus ({\bf 1,\,4,\,3})
\\({\bf 2,\,4,\,2}) \oplus ({\bf 2,\,4,\,2})
                                                \end{array} $\\\hline
({\bf 5, 8, 3}) & $\cU, \cV{}^{a\alpha}$ &  \p{5constr} &  $ \begin{array}{l}   ({\bf 1,\,4,\,3}) \oplus ({\bf 4,\,4,\,0})\\
                                                                  ({\bf 3,\,4,\,1}) \oplus ({\bf 2,\,4,\,2})
                                                \end{array} $\\\hline
({\bf 6, 8, 2}) & $\cV{}^{ij}, \cW{}^{ab}$ &  \p{7con1}, \p{7con2} &  $ \begin{array}{l}
({\bf 3,\,4,\,1}) \oplus ({\bf 3,\,4,\,1})\\
                                                                  ({\bf 4,\,4,\,0}) \oplus ({\bf 2,\,4,\,2})
                                                \end{array} $\\\hline
({\bf 7, 8, 1}) & $\cV{}^{ij}, \cQ{}^{a\alpha}$ &  \p{8con1}, \p{8con2} &
       $\begin{array}{l} ({\bf 3,\,4,\,1}) \oplus ({\bf 4,\,4,\,0}) \end{array}$ \\ \hline
({\bf 8, 8, 0}) & $\cQ{}^{aA}, \Phi{}^{i\alpha} $ &  \p{9con1}, \p{9con2} &
     $\begin{array}{l} ({\bf 4,\,4,\,0}) \oplus ({\bf 4,\,4,\,0}) \end{array}$\\
\hline
\end{tabular}
\end{center}
\vspace{0.5cm}

These results should be regarded as preparatory for a more detailed study
of the $N{=}8$ SQM models associated with the supermultiplets considered.
In particular, it would be interesting to reproduce these supermultiplets
{}from nonlinear realizations of $N{=}8, d{=}1$ superconformal groups,
construct the relevant superconformal actions and reveal
a possible relation of the corresponding $N{=}8$ SCQM to the physics of branes and black holes.
An intriguing question is whether some dynamical models with higher
$N{>}8, d{=}1$ supersymmetry can be constructed by combining some of the $N{=}8$
multiplets considered in this paper. For deeper geometrical understanding of the $N{=}8$
models proposed above, it would be interesting to consider their Hamiltonian formulation, as it
was done in \cite{BN} and \cite{BIGK} for $N{=}4$ SQM.

It should be pointed out that in the present work we addressed only those
multiplets which satisfy linear constraints in the superspace. As we know,
there exist $N{=}4, d{=}1$ multiplets with nonlinear defining constraints
(e.g. nonlinear versions of the chiral ({\bf 2, 4, 2}) multiplet \cite{IKL2},
as well as of the hypermultiplet ({\bf 4, 4, 0}) \cite{HP}).
It would be interesting to construct analogous nonlinear versions of some
$N{=}8$ multiplets from the above set. Also, in \cite{{IL},{IKL2}} plenty of nonlinear
off-shell relations between the $N{=}4,\, d{=}1$ superfields listed in Section~2 were found, as
a generalization of the linear relations described in
\cite{{GR},{PaTo}}. These substitutions preserve the number ${\bf 4}$ of the fermionic fields,
but express some auxiliary fields in terms of the time derivatives of physical bosonic
fields. In the $N{=}8$ case one can expect similar relations, and it is of interest to find
the explicit form of them and to explore their possible implications in $N{=}8$ SQM models.

Finally, in this paper we restricted ourselves to $N{=}8, d{=}1$ multiplets
with a {\it finite\/} number of off-shell components. It still remains to be
understood how they are related to multiplets with an infinite number of
auxiliary fields, which naturally appear in various versions of
{\it harmonic\/} $N{=}8, d{=}1$ superspace (see e.g. \cite{{BMZ},{BIK},{ISmi}}).
Also, the relevance of the latter to the $N{=}8$ SQM model building needs
to be explored further.

\section*{Acknowledgements}
This work is dedicated to the memory of our friend and colleague Tolya
Pashnev.

This research was partially supported by the European
Community's Human Potential
Programme under contract HPRN-CT-2000-00131 Quantum Spacetime,
the INTAS-00-00254 grant, the NATO Collaborative Linkage Grant PST.CLG.979389,
RFBR-DFG grant No 02-02-04002, grant DFG No 436 RUS 113/669, RFBR grant
No 03-02-17440 and a grant of the Heisenberg-Landau programme.
E.I. thanks the Institute of Theoretical Physics of the University
of Hannover and the Theory Group of the University of Padua
for the warm hospitality at the final stage of this work.
S.K. thanks INFN - Laboratori Nazionali di Frascati and the Institute of Theoretical Physics
of the University of Hannover for the warm hospitality extended to him during the course of this work.

\end{document}